\documentclass[twocolumn,showpacs,preprintnumbers,amsmath,amssymb,aps]{revtex4}
\usepackage{graphicx}
\usepackage{dcolumn}
\usepackage{bm}
\begin{document}
\preprint{}
\title{Axially-symmetric Neutron stars: Implication of rapid rotation}
\author{B. K. Sharma$^1$} \email{bksharma@tifr.res.in}
\author{T. K. Jha$^2$} \email{tkjha@prl.res.in}
\affiliation{$^1$Department of Nuclear and Atomic Physics, Tata Institute of
Fundamental Research, Homi Bhabha Road, Mumbai - 400 005, India.}
\affiliation{$^2$Theoretical Physics Division, Physical Research Laboratory,
Navrangpura, Ahmedabad - 380 009, India.}

\begin{abstract}
Models of relativistic rotating neutron star composed of hyperon rich matter is constructed in the framework of an effective field theory in the mean-field approach. The gross properties of compact star is calculated at both static and the mass-shedding limit in the axially symmetric basis. The effect of appearance and abundance of hyperons on equation of state of dense matter and stellar properties is lineated with particular emphasis on the underlying nuclear interactions. We find that the models can explain fast rotations, which supports the existence of millisecond pulsars. An important offshoot of the present investigation is that, irrespective of the model parameters and interaction taken, the star seems to sustain faster rotations (an increase in rotational frequency up to $\approx$ 50\%) without any further deformation.
\end{abstract}
\maketitle

\section{INTRODUCTION}

Neutron stars are the manifestation of the densest form of cold matter in the observable universe \cite{ba75,he00}. Formed in the aftermath of type II supernovae, they represent the end point of the life cycle of the star. Since all the known forces of nature amalgamate to determine the evolution and structural properties of these stars, they can provide vital link between astrophysics, nuclear physics, particle physics as well as heavy-ion collision data. Theoretically, the properties of neutron stars such as its mass, radius and related properties are the imprints of a particular equation of state (EoS). At $\approx (2-3) \rho_0$, where $\rho_0$ is the normal nuclear matter density, the presence of exotic forms of matter like hyperons, quark or mixed phase starts to appear and has substantial impact on the EoS, which in turn is reflected in the global properties of the star. Since we are still devoid of having any smoking gun signals from observations pertaining to the presence of these novel phase, apparently it becomes important to investigate and address these issues in a proper theoretical framework. It is known that the maximum mass of the star is regulated by the EoS well beyond 2$\rho_{0}$ and the radius is regulated by the EoS in the immediate vicinity of $\rho_{0}$. Arguably, in order to minimize the uncertainty in the mass and radius predictions, one needs to constrain the EoS both at $\rho_0$ and beyond. Till date, more than 1800 pulsars have been discovered with varying degrees of predictions, which invoke both challenge as well as interest in search for the realistic EoS for neutron stars. For example, on one hand, the evidence from the white dwarf-pulsar system J0751+1807 with a mass of M = $(2.1\pm0.2)M_{\odot}$ (1$\sigma$) and M = (1.6-2.5)$M_{\odot}$ (2$\sigma$) \cite{ni05} and a mass limit M$\geq(2.10\pm0.28)M_{\odot}$ from x-ray bursts of the neutron star EXO 0748-676 \cite{oz06} rule out the possibility of a soft EoS. On the other hand, the bounds from thermally emitting neutron stars eg. RXJ 1856.5-3754 \cite{wa02} imply the star radius to be beyond 10-12 km emphasize the need of a stiff EoS. Similarly, the subthreshold production of kaons as measured by the KaoS collaboration \cite{st01} were investigated in transport models \cite{fu01} which predicts the EoS to be soft. Also, recent results obtained from transport models reveals a soft symmetry energy at $\rho_B > 2\rho_0$ \cite{li09} has raised crucial implications on the nature of EoS. These apparent contradictions leaves ample opportunity to probe the various aspects of an EoS in detail and emphasize the need of constraints.

With this motivation, we now employ models of relativistic effective field theory \cite{mu96} to address these inevitable issues relating to the neutron star properties and discuss implications of fast rotations on the stellar structure. The model embodies various interactions and cross couplings, that satisfies the nuclear matter saturation properties and exploits the poorly known density dependence of the symmetry energy \cite{ho01,pi01}. The model is generalized to include the lowest lying octet of baryons and is employed to study the static as well as rotational attributes of the star.

The paper is organized as follows: In section II we give a brief description of the hadronic model which we use in our present work. In section III, we describe the numerical scheme to calculate the global properties of compact star. The main features of EoS and the global properties of the rotating neutron star is presented in section IV. We compare our results with other neutron star models in the light of observational bounds in section V. Finally, we conclude with the important findings of the present work in section VI.

\section{THE EQUATION OF STATE}

In the framework of Quantum hadrodynamics (QHD) for describing the nuclear many-body problem, as a relativistic systems of baryons and mesons, we employ an effective mean-field models to obtain the equation of state of hyperon rich matter up to densities relevant to the neutron star. The extra terms in the Lagrangian plays crucial role in determining the nature of an EoS, and is found to be in agreement with the Dirac-Bruckner-Hartree-Fock (DBHF) results at high density. The details of the hadronic model and the derivation of the equation of motion of the meson fields along with the energy density and pressure for nuclear matter can be found in Ref. \cite{sh07} and we shall only briefly outline the formalism here. The field equations for the baryons and mesons are obtained from the Lagrangian density which is given by,

\begin{eqnarray}
{\cal L} & = & \sum_{B}\overline{\Psi}_{B}\left ( i\gamma^\mu
D_{\mu} - m_{B} + g_{\sigma B}{\sigma}\right)
{\Psi}_{B}
+\frac{1}{2}{\partial_{\mu}}{\sigma}{\partial^{\mu}}{\sigma} \nonumber \\ 
&-& m_{\sigma}^2{\sigma^2}\left(\frac{1}{2}
+\frac{\kappa_3}{3 !}
\frac{g_{\sigma N}\sigma}{m_{N}}+\frac{\kappa_4}{4 !}
\frac{g^2_{\sigma N}\sigma^2}{m^2_{N}}\right)
-\frac{1}{4} {\Omega_{\mu\nu}}{\Omega^{\mu\nu}}
\nonumber \\
&+& \frac{1}{2}\left (1 +
{\eta_1}\frac{g_{\sigma N}\sigma}{m_{N}}
+\frac{\eta_2}{2}\frac{g^2_{\sigma N}\sigma^2}{m^2_{N}}\right)
m_{\omega}^2{\omega_{\mu}}
{\omega^{\mu}} - \frac{1}{4}{R^a_{\mu\nu}}{R^{a\mu\nu}} \nonumber \\
&+& \left(1 + \eta_{\rho}\frac{g_{\sigma N}\sigma}{m_{N}}\right)
\frac{1}{2}m_{\rho}^2{\rho^a_{\mu}}{\rho^{a\mu}}+\frac{1}{4 !}{\zeta_{0}}
g^2_{\omega N}\left({\omega_{\mu}}{\omega^{\mu}}\right)^2
\end{eqnarray}

In the above lagrangian the covariant derivative ${D_{\mu}}$ is defined as ${D_{\mu}} = \partial_{\mu} + ig_{\omega B}{\omega_{\mu}} + ig_{\rho B}I_{3B}{\tau^a}{\rho^a_{\mu}}$, where $R^a_{\mu\nu}$ and $\Omega_{\mu\nu}$ are the field terms for the vector ($\omega$) and the isovector meson ($\rho$) defined as follows.

\begin{eqnarray}
{\Omega_{\mu\nu}} & = & \partial_{\mu}\omega_{\nu} - \partial_{\nu}\omega_{\mu}.
\\
R^a_{\mu\nu} & = & \partial_{\mu}\rho^a_{\nu} - \partial_{\nu}\rho^a_{\mu}
+ g_{\rho}\epsilon_{abc}\rho^b_{\mu}\rho^c_{\nu}.
\end{eqnarray}

The above lagrangian (eqn. 1) embodies scalar meson interaction terms up to quartic order with $\kappa_3$ and $\kappa_4$ as the higher order scalar field constants for the cubic and the quartic terms respectively. Similarly for the vector ($\omega-$meson) interactions, in addition to the usual second and fourth order mass terms, the scalar-vector cross couplings are also taken, where $\eta_1$ and $\eta_2$ are the corresponding coupling constants. Similarly, the iso-scalar and iso-vector cross interaction term is also included (last line), where $\eta_{\rho}$ is the constant that enters as a parameter in the model. Finally $\zeta_0$ is the constant for the quartic vector interaction. Therefore, in addition to the scalar meson mass, and the three coupling constants namely, $g_{\sigma}$,$g_{\omega}$ and $g_{\rho}$, there are six other parameters that is tuned to reproduce the nuclear matter saturation properties. The influence of the inclusion of extra terms in the lagrangian on nuclear saturation properties can be seen in Table 1, where we have specified the models along with the corresponding saturation properties given in the lower panel. 

The meson fields equations for $\sigma$, $\omega$ and $\rho$-meson, in the mean-field ansatz are given by

\begin{eqnarray}
&& m^2_\sigma\left(\sigma_0 + \frac{g_{\sigma 
N}\kappa_{3}}{2m_{N}}{\sigma^2_0} +\frac{g^2_{\sigma
N}\kappa_{4}}{6m_{N}^2}{\sigma^3_0}\right) \nonumber \\
&-& \frac{1}{2}m^2_{\omega}
\left(\eta_1\frac{g_{\sigma N}}{m_{N}} + \eta_2\frac{g^2_{\sigma
N}}{m_{N}^2}\sigma_0\right) {\omega_0^2} \nonumber \\
&-& \frac{1}{2}m^2_{\rho}\eta_{\rho}\frac{g_{\sigma N}}{m_{N}}\rho^2_0 =
\sum_{B}g_{\sigma B}{m^*}^2_{B}\rho_{SB},
\end{eqnarray}
\begin{eqnarray}
& & m^2_{\omega}\left(1 + \frac{{\eta_1}g_{\sigma N}}{m_{N}}\sigma_0 +
\frac{{\eta_2}g^2_{\sigma N}}{2m_{N}^2}\sigma^2_0\right){\omega}_0 \nonumber \\
&+& \frac{1}{6}{\zeta_0} g^2_{\omega N}{\omega^3_0} = \sum_{B}g_{\omega B}\rho_B,
\end{eqnarray}

\begin{eqnarray}
& & m^2_{\rho}\left(1 + \frac{g_{\sigma
N}{\eta_{\rho}}}{m_{N}}\sigma_0\right)\rho_{03} = \sum_{B}g_{\rho
B}I_{3B}\rho_B.
\end{eqnarray}
The scalar $(\rho_{SB})$ and the vector densities $(\rho_B)$, for a particular baryon species  are given by,
\begin{equation}
\rho_{SB}= \frac{1}{\pi^2}\int_{0}^{k_B} \frac{k^2
dk}{E^*_B},
\end{equation}
\begin{equation}
\rho_{B} = \frac{1}{\pi^2}\int_{0}^{k_B} {k^2 dk},
\end{equation}
where $E^{*}_{B}=\sqrt{k^{2}+{m^*}^2_{B}}$ is the effective energy and $I_{3B}$ is the isospin projection of $B$, the quantity $k_B$ is the Fermi momentum for the baryon. The effective mass for the baryon species is then given by $m^*_{B}=m_B-g_{\sigma B}\sigma$.

The meson field equations for the $\sigma$, $\omega$ and $\rho-$mesons are then solved self-consistently at a fixed baryon density to obtain the respective field strengths. The EoS for the $\beta-$equilibrated for the hyperon rich matter is obtained with the requirements of conservation of total baryon number and charge neutrality condition given by,

\begin{equation}
\sum_{B}Q_{B}\rho_{B}+\sum_{l}Q_{l}\rho_{l}=0,
\end{equation} 
\noindent
where $\rho_{B}$ and $\rho_{l}$ are the baryon and the lepton (e,$\mu$) number densities with $Q_{B}$ and $Q_{l}$ as their respective electric charges. The lepton Fermi momenta are the positive real solutions of $(k_e^2 + m_e^2)^{1/2} =  \mu_e$ and $(k_\mu^2 + m_\mu^2)^{1/2} = \mu_\mu = \mu_e$.

The corresponding pressure and energy density of the charge neutral beta-equilibrated neutron star matter(which includes the lowest lying octet of baryons) is then given by

\begin{eqnarray}
P &=& \sum_{B} \frac{2}{3(2\pi )^{3}}\int_{0}^{k_B}
d^{3}k\frac{k^{2}}{E^{*}_{B}(k)} 
+ \frac{1}{ 4!}\zeta _{0}g_{\omega N}
^{2}{\omega_{0}}^{4}
\nonumber\\ 
&&+ \frac{1}{2}\left(1+\eta _{1}\frac{g_{\sigma N}\sigma_{0}}
{m_{N}} 
+ \frac{\eta _{2}}{2}\frac{g_{\sigma N}^{2}\sigma_{0}^{2}}{m_{N}^{2}}\right)
m_{\omega}^{2}{\omega_{0}^{2}}
\nonumber \\
&& -m_{\sigma}^{2}\sigma_{0}^{2}\left(\frac{1}{2}+\frac{\kappa_{3}g_{\sigma N}
\sigma_{0}}{3!m_{N}} +\frac{\kappa_{4}g_{\sigma N}^{2}\sigma_{0}^{2}}{4!m_{N}^{2}}
\right)
\nonumber\\
&& + \frac{1}{2}\left(1+\eta
_{\rho }\frac{g_{\sigma N}\sigma_{0}}{m_{N}}\right)m_{\rho
}^{2}\rho_{0}^{2} +\sum_{l}P_{l}\;,
\end{eqnarray}

\begin{eqnarray}
{\cal{E}} &=& \sum_{B} \frac{2}{(2\pi )^{3}}\int_{0}^{k_B}
d^{3}k{E^{*}_{B}(k)} + \frac{1}{8}\zeta _{0}g_{\omega N}
^{2}{\omega_{0}}^{4} 
\nonumber\\
&&+ \frac{1}{2}\left(1+\eta _{1}\frac{g_{\sigma N}\sigma_{0}}
{m_{N}} + \frac{\eta _{2}}{2}\frac{g_{\sigma N}^{2}\sigma_{0}^{2}}{m_{N}^{2}}\right)
m_{\omega}^{2}{\omega_{0}^{2}}
\nonumber \\
&& +m_{\sigma}^{2}\sigma_{0}^{2}\left(\frac{1}{2}+\frac{\kappa_{3}g_{\sigma N}
\sigma_{0}}{3!m_{N}} +\frac{\kappa_{4}g_{\sigma N}^{2}\sigma_{0}^{2}}{4!m_{N}^{2}}
\right) 
\nonumber\\
&& + \frac{1}{2}\left(1+\eta
_{\rho }\frac{g_{\sigma N}\sigma_{0}}{m_{N}}\right)m_{\rho
}^{2}\rho_{0}^{2} +\sum_{l}{\cal{E}}_{l}.
\end{eqnarray}

The subscript `$B$' in the above equations corresponds to the lowest lying octet of baryons and `$N$' and `$l$' specifies the nucleon and leptons respectively. Here $P_{l}$ and $\varepsilon_{l}$ are the contributions to the total pressure and energy density respectively from the leptonic counterparts. The corresponding model parameters that enters in our calculation of the equation of state at specified in Table I.

\begin{table}[h]
\caption{The model parameters from the relativistic mean-field theory. The lower panel display the nuclear matter saturation properties.}
\centering
\begin{tabular}{llllcccccccccccccccccccccrrrrr}
\\
\hline
\hline
\multicolumn{1}{c}{ } &&
\multicolumn{1}{c}{G2} &&
\multicolumn{1}{c}{G1} &&
\multicolumn{1}{c}{TM1*} &&
\multicolumn{1}{c}{TM1} &&
\multicolumn{1}{c}{NL3} &&
\\
\hline
$m_{s}/m_{N}$       &&  0.554 &&  0.540  && 0.545 && 0.545 && 0.541  \\
$g_{sN}/4{\pi}$     &&  0.835 &&  0.785  && 0.893 && 0.798 && 0.813  \\
$g_{vN}/4{\pi}$     &&  1.016 &&  0.965  && 1.192 && 1.003 && 1.024  \\
$g_{\rho N}/4{\pi}$ &&  0.755 &&  0.698  && 0.796 && 0.737 && 0.712  \\
$\kappa_{3}$        &&  3.247 &&  2.207  && 2.513 && 1.021 && 1.465  \\
$\kappa_{4}$        &&  0.632 && -10.090 && 8.970 && 0.124 && -5.668 \\
$\zeta_{0}$         &&  2.642 &&  3.525  && 3.600 && 2.689 &&  0.0  \\
$\eta_{1}$          &&  0.650 &&  0.071  && 1.1   && 0.0   && 0.0  \\
$\eta_{2}$          &&  0.110 && -0.962  && 0.1   && 0.0   && 0.0  \\
$\eta_{\rho}$       &&  0.390 && -0.272  && 0.45  && 0.0   && 0.0  \\
\hline
$a_{v} (MeV)$       && -16.07 &&  -16.14 && -16.30 && -16.30 && -16.24  \\
$\rho_{0} (fm^{-3})$&&  0.153 &&   0.153 && 0.145  && 0.145  && 0.148   \\
$K$ (MeV)           &&   215  &&   215   && 281.1  && 281.1  && 271.5   \\
$M^*_N/M_N$         &&  0.664 && 0.634   && 0.634  && 0.634  && 0.595   \\
$J$ (MeV)           &&  36.4  &&   38.5  && 36.90  && 36.90  && 37.40   \\
\hline
\hline
\end{tabular}
\end{table}

From the table, one can clearly outline the differences in the nuclear matter saturation properties, which is attributed to the underlying interactions. For example, model $G1$ and $G2$, represents a class of model which entertains all the interactions described in the lagrangian and hence has the lowest value of incompressibility in comparison to other mean-field models given in Table I, i.e., the resulting EoS is softer. However, it is interesting to find that, although TM1* has the same attributes, yet the resulting incompressibility and all other saturation properties remains the same as TM1, where the former is a modification of the later with new couplings which stem from the modern effective field theory approach to relativistic nuclear phenomenology. Among all the present models, NL3 has the lowest value of effective nucleon mass. It is to be noted that the $\kappa_{4}$ have a positive sign in TM1, TM1* and G2 unlike NL3 and G1. The positive sign of $\kappa_{4}$ is required so that energy spectrum has a lower bound, else instabilities in calculations of the equation of state and of finite systems may occur \cite{baym60}. In case of TM1* the inclusion of new couplings not only give the results close to DBHF at high density, it is also applicable to finite nuclei calculation for $Z < 8$ \cite{patra}. It would be interesting to analyze the effect of these fundamental differences at high density and neutron star properties.

\section{Stellar Equations}

The EoS of dense matter is the primary ingredient required to evaluate the global properties of the neutron stars. In order to account for the crustal part of the compact star, we include the BPS equation of state \cite{bps} initially followed by the charge neutral, beta equilibrated hyperon rich matter, as described in the previous section. However, in addition one needs to specify the hyperon coupling strengths . Presently, we assume that all the hyperons have the same couplings, which are in accordance with the quark sum rule approach. Accordingly, the scalar, vector and the iso-vector coupling strength for the hyperons are are expressed in terms of the nucleon coupling as $x_{\sigma}=x_{H\sigma}/x_{N\sigma}=\sqrt{2/3}$, $x_{\omega}=x_{H\omega}/x_{N\omega}=\sqrt{2/3}$ and $x_{\rho}=x_{H\rho}/x_{N\rho}=\sqrt{2/3}$. The equations for the structure of a relativistic spherical and static star composed of a perfect fluid were derived from Einstein's equations by Tolman, Oppenheimer and Volkoff \cite{tov}, which are

\begin{equation}
\frac{dP}{dr}=-\frac{G}{r}\frac{\left[\varepsilon+P\right ]
\left[M+4\pi r^3 P\right ]}{(r-2 GM)},
\label{tov1}
\end{equation}
\begin{equation}
\frac{dM}{dr}= 4\pi r^2 \varepsilon,
\label{tov2}
\end{equation}
\noindent

with $G$ as the gravitational constant and $M(r)$ as the enclosed gravitational mass. We have used $c=1$. Given an EoS, these equations can be integrated from the origin as an initial value problem for a given choice of central energy density, $(\varepsilon_c)$. The value of $r~(=R)$, where the pressure vanishes defines the surface of the star. We solve the above equations to study the structural properties of a static neutron star using the EoS for the electrically charge
neutral hyperonic dense matter \cite{nr,nr1}. In case of an evolved neutron star, the approximation of a zero temperature perfect fluid neutron star matter is reasonable \cite{fluid}. Assuming a uniform rotation with static, axial symmetric space-time, the time translational 
invariant and axial-rotational invariant metric in spherical polar coordinates 
(\textit{t,r}, $\theta,\phi$) can be written as

\begin{equation} 
ds^2=-e^{2\nu} dt^2+e^{2\alpha} (dr^2+r^2 d\theta^2) 
+e^{2\beta}r^2\sin^2\theta (d\phi - \omega dt)^2, 
\label{metric}
\end{equation}
\noindent
where the metric functions $\nu, \alpha, \beta, \omega$ depends only on $\textit{r}$ and $\theta$. For a perfect fluid, the energy momentum tensor can be given by

\begin{equation}
T^{\mu\nu} = Pg^{\mu\nu} + (P+\epsilon) u^{\mu} u^{\nu},
\end{equation}
\noindent
with the four-velocity 

\begin{equation}
u^{\mu} = \frac{e^{-\nu}}{\sqrt{1-v^2}}(1,0,0,\Omega).
\end{equation}
\noindent
Here 

\begin{equation}
\mathit{v}=(\Omega-\omega) r\sin\theta e^{\beta-\nu}, \label{vsin}
\end{equation}
\noindent
is the proper velocity relative to an observer with zero angular velocity and $\Omega$ is the angular velocity of the star measured from infinity. Now we can compute the Einstein field equations given by

\begin{eqnarray}
\mathit{R}_{\mu \nu} - \frac{1}{2} g_{\mu \nu}\mathit{R} = {8\pi}T_{\mu \nu}
\end{eqnarray}
\noindent
(where $\mathit{R}_{\mu \nu}$ is the Ricci tensor and and $\mathit{R}$ is the 
scalar curvature and with c = G = 1). A maximum limit for the stable rotation of a star $\Omega_{K}$, is set by the onset of mass shedding from the equator of the star. General relativistic expression for this Keplerian frequency `$\Omega_{K}$' can be obtained by using the extremal principle to the circular orbit of a point particle rotating at the equator of the star \cite{NKGFB}. We use the code written by Stergioulas \cite{steir} based on the Komatsu-Eriguchi-Hachisu method to construct uniformly rotating star models.

\section{RESULTS AND DISCUSSION}

The extreme of matter density and high pressure prevailing in the core of neutron star pave  way for exotics such as hyperons or quarks. In the present work, however we are concerned only with the aspects and possible implications of the presence of hyperons (lowest lying octet of baryons) in the core of these stars. The ambiguity or lack of precise experimental data for the hyperon-hyperon or hyperon-meson interaction compels us to take certain assumption. Therefore, presently we take the hyperon-meson couplings in accordance with the quark sum rule approach as described in the previous section and is taken to be same for all the hyperon species. Although few hypernuclei experiments exist, but only for the case of $\Lambda^0$, which shows bound state of the particle at normal nuclear density, which comes out to be $\approx$ - 30 MeV. Theoretically, Hyperons are known to be stable and dominant species in the matter composition at high pressure and dense environment. Their appearance and abundance in the matter is known to have substantial contribution to the energy of the system rather than the pressure, which gives rise to a net softening effect on the EoS. Quantitatively, the amount of softening depends on factors, such as the underlying model attributes at saturation density and on the choice of coupling taken. However, previous studies with hyperons in dense matter and neutron stars seems to agree on two important aspects, each of which has important implications to the neutron star structure and composition. One of which is the appearance and abundance of hyperon species and the second one is the resulting deleptonization of dense matter. Leptons primarily gets used up to maintain the charge neutrality of the matter leading to the deleptonization, which does not happen in the case of nucleon matter composition. To account for the crustal effect, we included the BPS EoS at subnuclear densities to the EoS of the hyperon rich neutron star matter. 

The resulting $\beta-$equilibrated EoS for hyperon rich matter is shown in Fig. 1(A). From the plot, it can be seen that NL3 which has the lowest nucleon effective mass predicts stiff EoS than the rest of the models. Although not much of a difference is seen in the energy per nucleon as well as saturation density for the models , there is substantial difference in the incompressibility and nucleon effective mass on account of its sensitivity to the EoS. However, beyond the normal nuclear matter density, the EoS is sensitive to the appearance and abundance of the hyperon species. Fig. 1(B) shows the density at which the first two members of the lowest lying octet of barons ($\Lambda^0$(1116) and $\Sigma^-$(1193)) start appearing in dense matter. The plot reveals that although $\Sigma^-$ is $\approx~77$ MeV massive than $\Lambda^0$, still they are the first ones to appear in dense matter owing to their negative isospin. $\Sigma^-$ seems to start appearing at nearly $(1.5 - 1.7) \rho_0$ closely followed by $\Lambda^0$ at $(1.7 - 2.2) \rho_0$. The resulting softening effect on the EoS is reflected on the EoS for the models at $\varepsilon ~\approx~(200 - 250) MeV fm^{-3}$. The successive appearance of other hyperon species adds to the softening of the equation of state thereafter. 

\begin{figure}[ht]
\begin{center}
\includegraphics[width=5.5cm,angle=0]{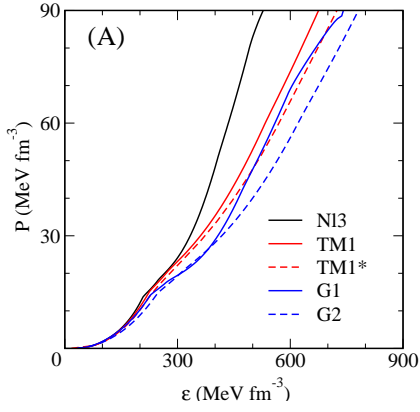}
\vskip 0.35in
\includegraphics[width=5.5cm,angle=0]{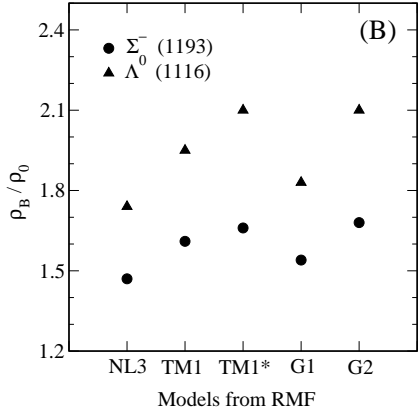}
\end{center}
\caption
{(Color online)(A): Equation of state ($\varepsilon$~ vs.~p) of hyperon rich neutron star matter for the five models from the relativistic mean-field theory. (B): Normalised baryon densities ($\rho_B / \rho_0$) at which $\Sigma^-$ and $\Lambda^0$ appears in dense matter, for the corresponding models.}
\end{figure}

\begin{figure}[ht]
\begin{center}
\includegraphics[width=5.5cm,angle=0]{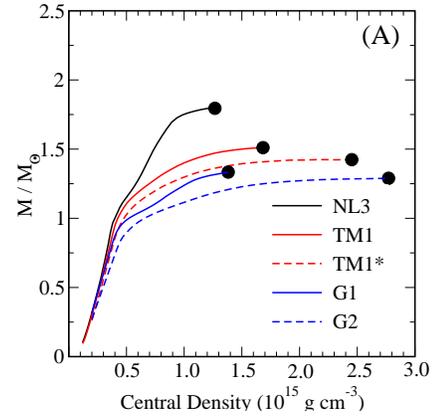}
\vskip 0.32in
\includegraphics[width=5.9cm,angle=0]{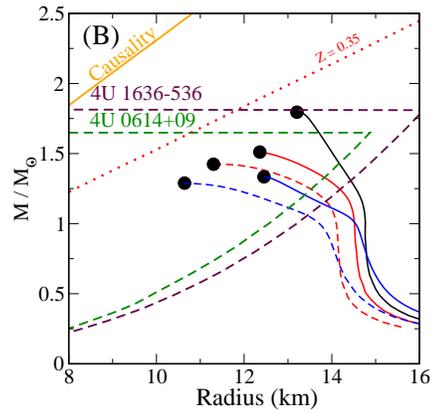}
\end{center}
\caption
{(Color online) For static case (A): Mass of the neutron star sequence plotted as a function of the central density of the star. (B): Mass of the neutron star as a function of radius is shown along with the constraints from neutron star observation and causality condition.}
\end{figure}

With the five EoS from the relativistic mean-field models as input, we now calculate the corresponding stellar properties for static case. Fig. 2(A), shows the mass of the neutron star obtained as a function of the central density of the star in the static limit. The filled circles denote the point of maximum mass for the corresponding models, presently under investigation. The quantitative impact of the onset of hyperons on the EoS and hence on the neutron star mass can be known if we compare the densities at which they appear. For example, `$\Sigma^-$' and `$\Lambda^0$' starts appearing nearly at similar density in case of TM1* and G2 and hence the resulting EoS follows similar patters although G2 predicts the softest EoS among the five models considered here. Because of the similar behavior and softening at high density, they also result in star with high central density ($\varepsilon_c$). For example, in case of TM1*, $\varepsilon_c \approx 9 \rho_0$ and for G2, it is $\approx$ $10\rho_0$. For the present models, the central density of the star at maximum mass lies in the range ($4-10$) $\rho_0$. The corresponding sequence of neutron stars in the M-R plane for static case is shown in Fig. 2(B) along with available constraints. The maximum mass of the static star lies in the range ($1.29 - 1.79$) $M_{\odot}$ and the corresponding compactness ratio (M/R) is in the range ($0.11 - 0.14$). It is interesting to see that, the results compare very well with the observation of massive stars $(M \approx 2 M_{\odot})$ \cite{mass,mass1} from observation of QPO's (Quasi Periodic Oscillations) from X-ray emmisions. However, there are also observation of least massive stars with mass $1.18 \pm 0.02 M_{\odot}$ \cite{least} from the binary pulsar $J1756-2251$. These lower and the upper bound on mass from observations have raised interesting implications on their structure and composition \cite{baym}. However, the canonical value of $M = 1.44 M_{\odot}$ still remains the largest precisely measured mass of the Neutron star $PSR 1913+16$ \cite{exact}. Under these circumstances, precise radius measurement holds the key to understand the structural properties of these stars. From the observational point of view, there are large uncertainties in determination of the radius of the star \cite{rad1,rad2,cott02} which is primarily because of our lack of knowledge of the composition of the star atmosphere, large distance and also due to the presence of high magnetic fields.However, it is worth to recall that although there are large observational errors in the mass-radius determination, yet in case of $Vela X-1$ the lower mass limit $\approx (1.6-1.7)M_{\odot}$ is at least mildly constrained by geometry \cite{geo}. Theoretically the mass of these compact stars are controlled by the nature of the EoS (stiffness/ softness) at an excess of $5 \rho_0$, whereas the determination of radius is controlled by the EoS in the vicinity of $\rho_0$, the nuclear matter saturation density. The region excluded by the causality constraint i.e., $\sqrt{\partial p/ \partial \varepsilon}~\le~1$ leads to $R > 2.9 GM/c^2$ is shown in Fig. 2(B) along with the redshift measurement from active X-ray burster EXO 0748-676, which implies $z \simeq 0.35$ \cite{cott02}. Although this value may not be universally accepted, yet EXO 0748-676 seems to be a promising candidate as it can provide stringent constraint in the M-R plane. The experimental constraints are extracted from observations of kilohertz quasi-periodic brightness oscillations in the Low Mass X-Binaries 4U 0614+09 and 4U 1636-536. Here we find that G2 results in low mass with highest central density owing to the soft EoS due to the underlying cross couplings included in the interactions. 

\begin{figure}[ht]
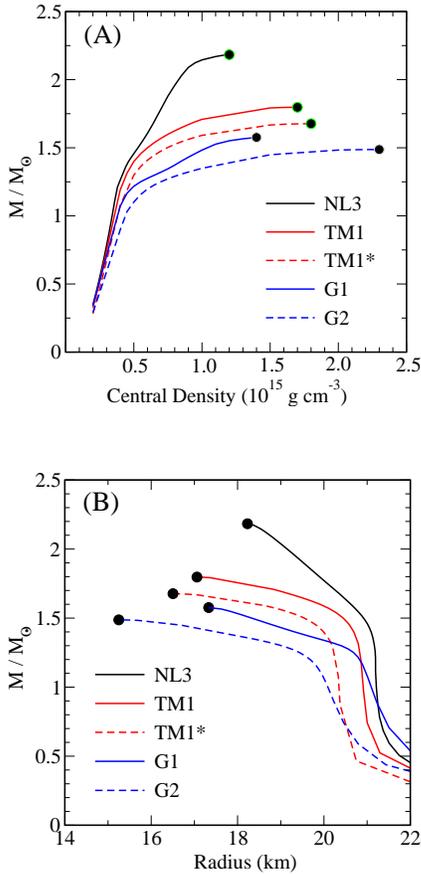

\begin{center}
\vskip 0.2in
\includegraphics[width=5.5cm,angle=0]{cd-m01-kep.eps}
\vskip 0.32in
\includegraphics[width=5.5cm,angle=0]{m-r01-kep.eps}
\end{center}
\caption
{(Color online) For star rotating at kepler velocity (A): Mass of the neutron star sequence plotted as a function of the central density of the star. (B): Mass as a function of radius for the corresponding sequence. The filled circles denotes the values at maximum mass.}
\end{figure}

\begin{figure}[ht]
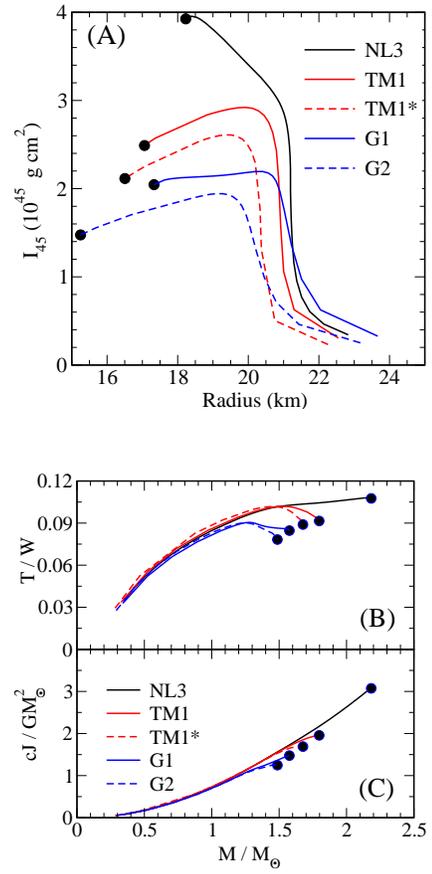

\begin{center}
\includegraphics[width=5.5cm,angle=0]{i-r01-kep.eps}
\vskip 0.3in
\includegraphics[width=5.5cm,angle=0]{m-t-j.eps}
\end{center}
\caption
{(Color online) For star rotating at kepler velocity (A): Variation of the Moment of Inertia `$I_{45}$ ($10^{45} g cm^2$)' of the neutron star sequence as a function of the radius of the star. (B): The corresponding gravitational energy of the star as a function of mass is shown. (C) The angular momentum of the star is shown as a function of mass for the five model under consideration. The filled circles denotes the values at maximum mass.}
\end{figure}

\begin{figure}[ht]
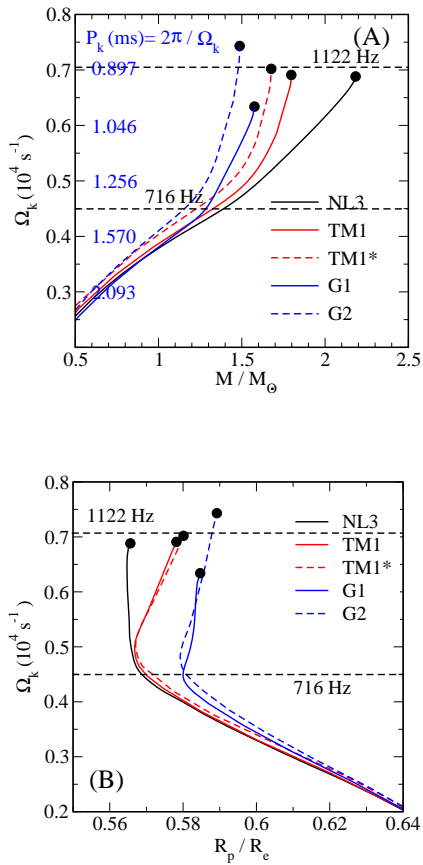

\begin{center}
\includegraphics[width=5.5cm,angle=0]{m-w01-kep.eps}
\vskip 0.4in
\includegraphics[width=5.5cm,angle=0]{rp-re-w.eps}
\end{center}
\caption
{(Color online) (A): The kepler velocity of the star is plotted as a function of mass of the star. (B): The kepler velocity of the star as a function of the resulting deformation parameter ($R_p / R_e$). The filled circles denotes the values at maximum mass. The rotational frequency of the two fastest known pulsars, rotating at $1122 Hz$ and $716 Hz$ are also indicated (dashed horizontal lines).}
\end{figure}

We now calculate the global properties of the neutron star undergoing keplerian rotation. Compared to static case, the effect of rotation is to increase the equatorial radius of the star and also to increase the mass that can be sustained at a given central energy density. For star rotating at kepler velocity we show the obtained mass of the neutron star sequence plotted as a function of the central density of the star in Fig. 3(A). We find that the mass of the neutron star increases for all the models, however the resulting central density of the star drops. Quantitively, we find an increase of $\approx$ ($15 - 20$) \% in the stellar mass and the decrease in the central density is $\approx$ ($5 - 17$) \% which corresponds to $\approx (0.2 - 1.6) \rho_0$. At the mass-shedding limit, the maximum mass of the star is found in the range ($1.49 - 2.18$) $M_{\odot}$. Fig. 3(B) shows the corresponding models of neutron stars at $\Omega = \Omega_k$ (mass-shedding limit) in the M-R plane. The filled circles shows the point of maximum mass for the models. In comparison to the static star, we find an increase of $\approx (38 - 43)$\% in the equatorial radius at maximum mass. The increase in the mass or equatorial radius are EoS dependent and in turn sensitive to the underlying model attributes at $\rho_0$. For example, in the framework of an effective chiral model, the neutron star properties in the kepler limit was studied and there an increase of nearly 50\% was found in the star radius \cite{tkj08}. This was a consequence of the underlying softness of the EoS. Similar correlations can be found in other related properties of the rapid rotors, which we discuss next.

Fig. 4(A) display the moment of inertia of the sequence of rapid rotors as a function of the radius of the star. Moment of inertia is a sensitive quantity for rapid rotors in addition to the fact that its accurate determination from observations would imply a higher degree of accuracy in the determination of neutron star radius, which still remains a major hurdle in our present understanding of these stars. Dimensionally, $I~\propto~MR^2$ and they play crucial role in the models of radio pulsar. Independent of the rotation, i.e., slow or fast, the relation of moment of inertia to the matter distribution within the star is complicated. For the models presently under investigation, the moment of inertia of the compact stars lies in the range $I_{45}$ = ($1.5 - 3.9$), where G2 results in lower value and NL3 has the highest value and has the largest compactness parameter (M/R) among them. The general relativistic limit for the compactness parameter assuming a uniform density star with the causal equation of state i.e., $P=\varepsilon$ gives $M/R < 4/9$ \cite{mr}. The compactness ratio for the present models lies far below the aforesaid limit and is found in the range $M/R = (0.10 -0.12)$. From the observational point of view, recently discovered relativistic pulsar $PSR J0737-3039$ \cite{mi-pulsar} could be the first one with moment of inertia measured precisely. However, the estimates from crab pulsar \cite{crab} puts $I_{45} > 1.61$ for $M_{neb}=2.0 M_{\odot}$, which is considered to be the conservative estimate, and $I_{45} > 3.04$ for $M_{neb}=4.6 M_{\odot}$ (latest estimate). Fig. 4(B) and Fig. 4(C) shows the variation of the gravitational/ rotational energy of the star rotating at kepler velocity as a function of stellar mass and the corresponding angular momentum of the rapid rotors respectively. As anticipated NL3 stands apart from the rest of the models, which reflects the stiffness of the corresponding EoS.

\begin{figure}[htbp]
\vskip 0.7in
\begin{center}
\includegraphics[width=3cm,angle=-90]{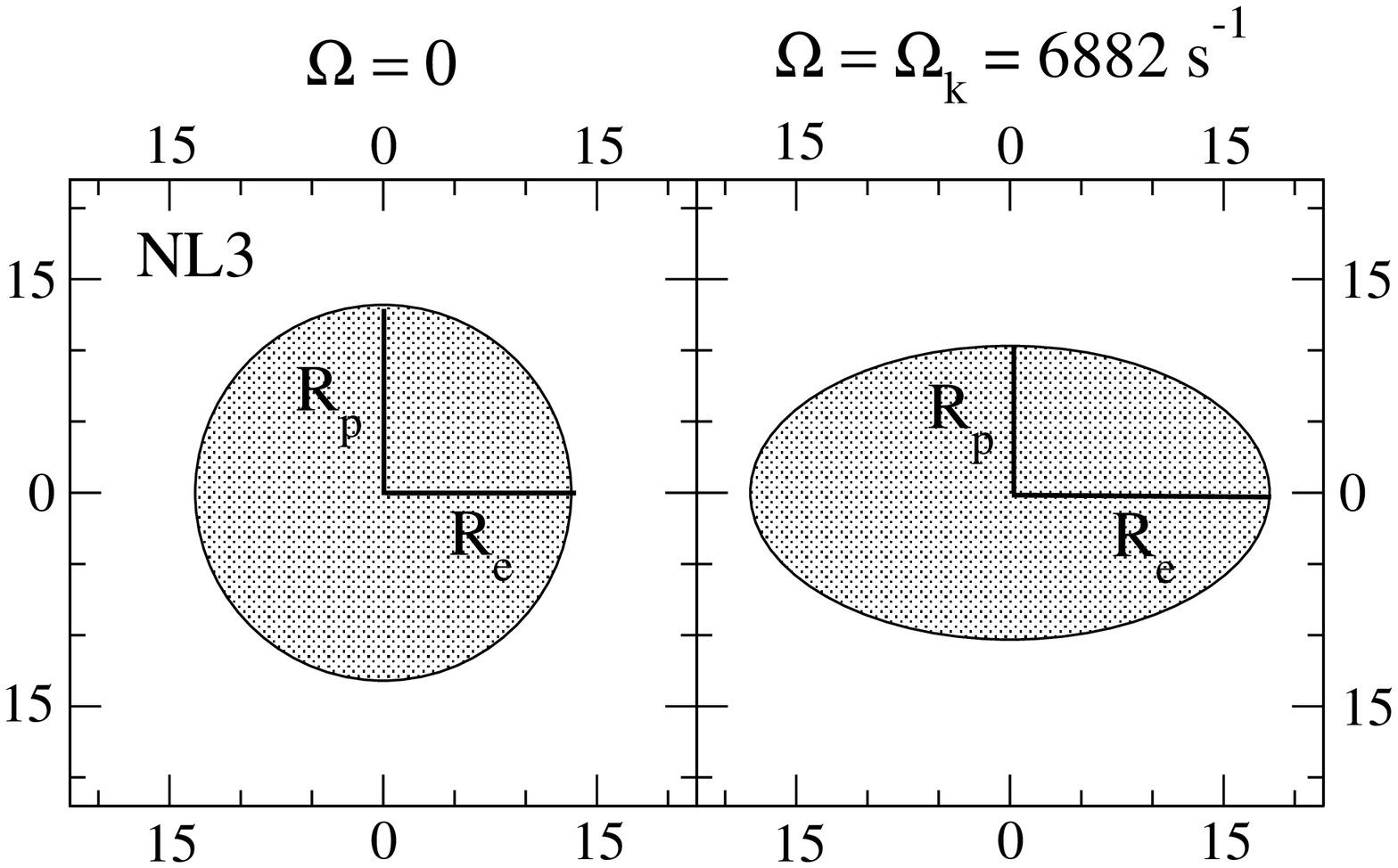}
\vskip 0.35in
\includegraphics[width=3cm,angle=-90]{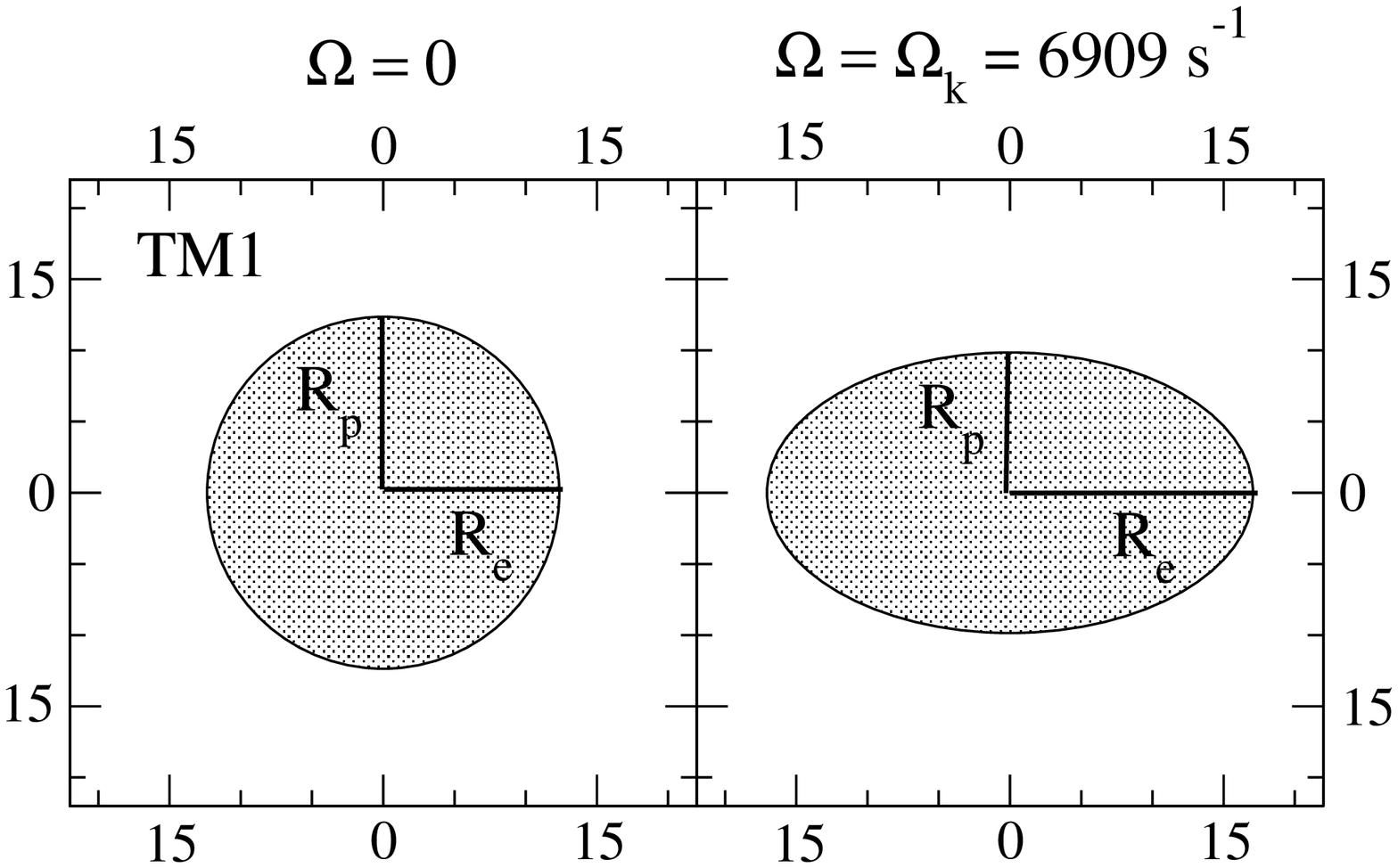}
\vskip 0.35in
\includegraphics[width=3cm,angle=-90]{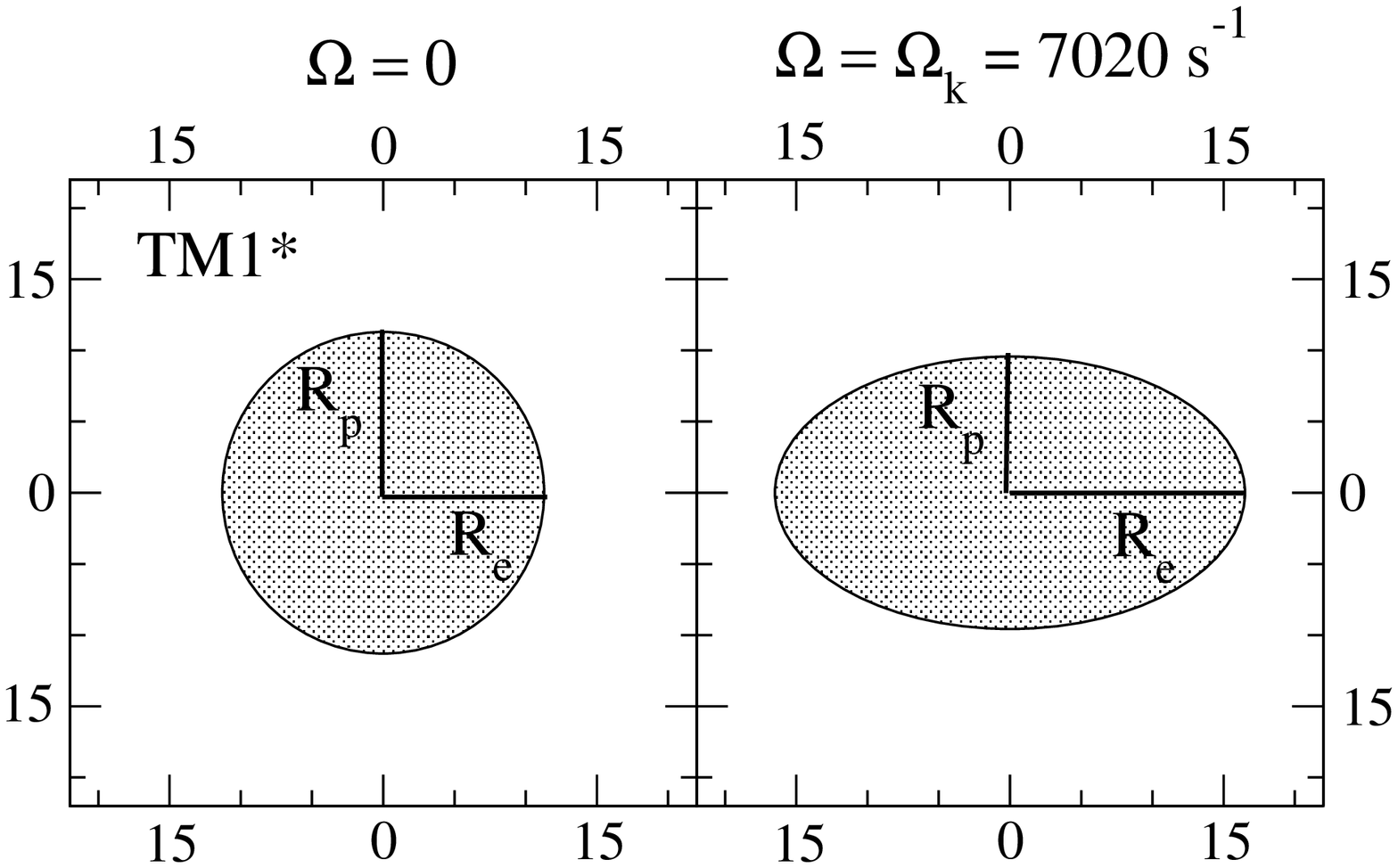}
\vskip 0.35in
\includegraphics[width=3cm,angle=-90]{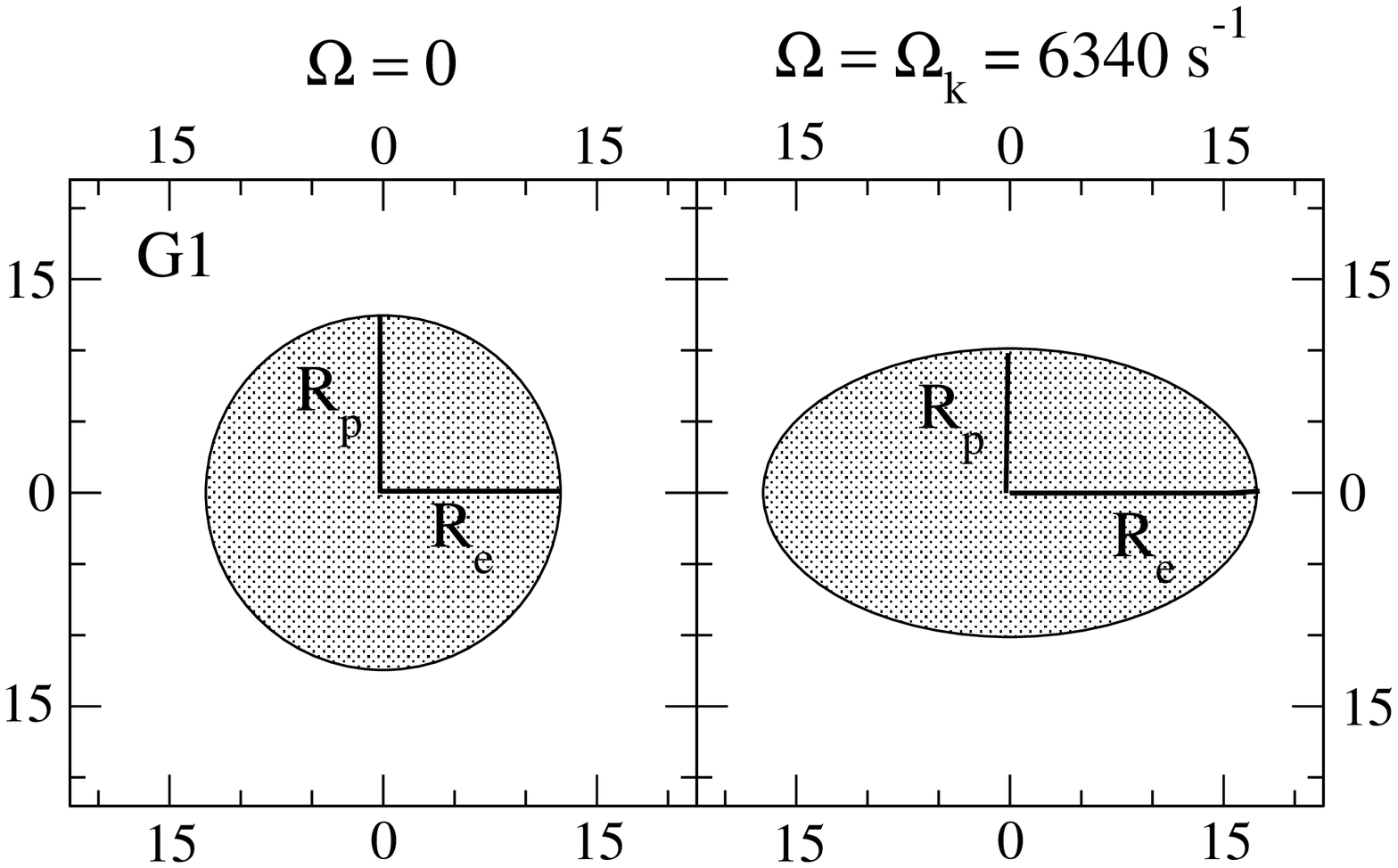}
\vskip 0.35in
\includegraphics[width=3cm,angle=-90]{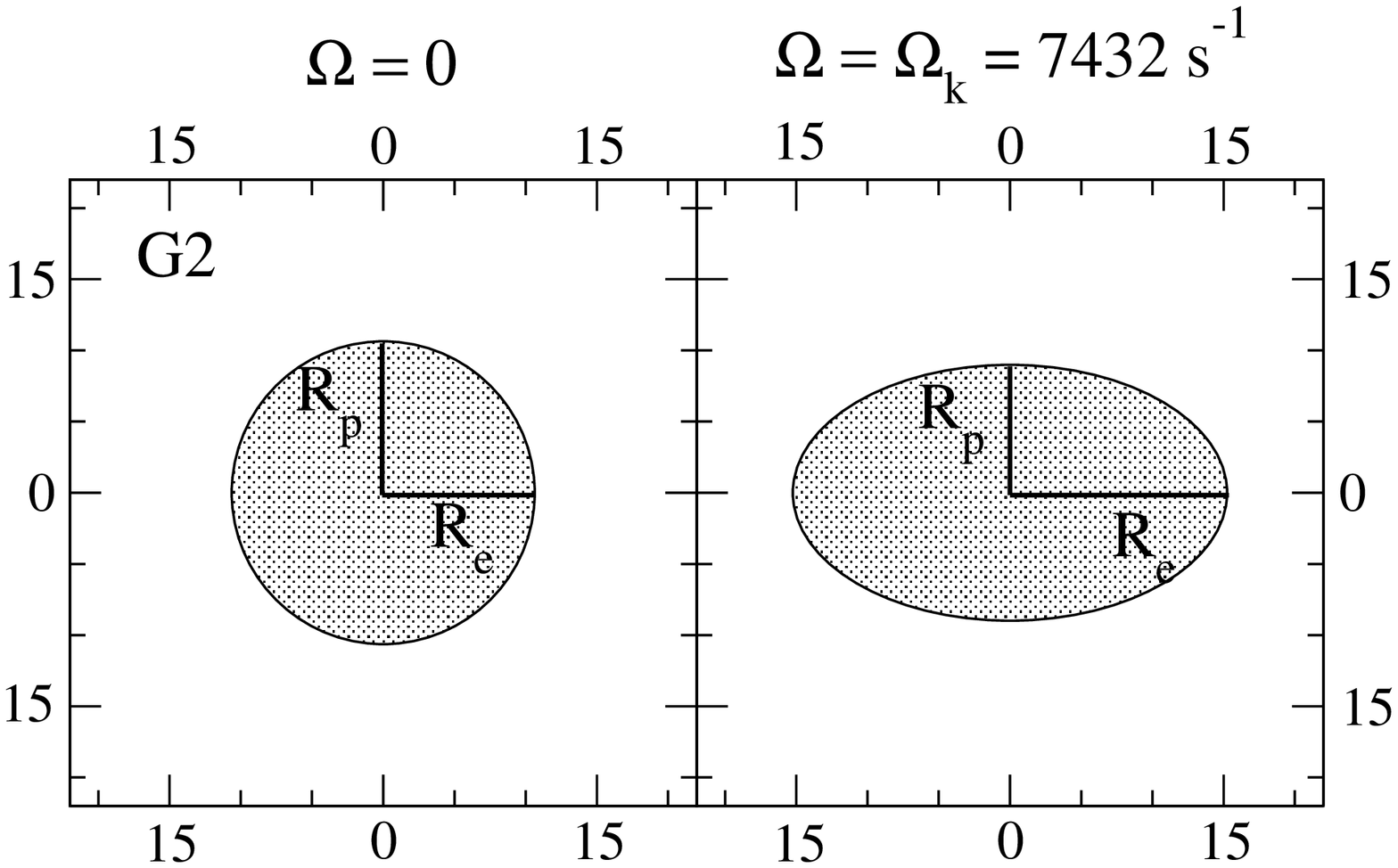}
\end{center}
\vskip 0.2in
\caption
{(Color online) The stellar structure for the static ($\Omega = 0$; left one) and the corresponding ones rotating at kepler velocity ($\Omega = \Omega_k$; right one) is shown for the models studied in the coordinates of polar radius (y axis) and the equatorial radius (x axis) in km.}
\end{figure}

Fig. 5(A) shows the variation of kepler velocity of the star as a function of star mass. Initially, we find that the rotation of the star increases linearly with mass for all the cases, but after the star attains a maximum mass of $\approx (1.4-1.5) M_{\odot}$, they undergo much rapid rotation but the increase in mass is not so substantial with an exception of NL3. In the inset on y axis, the corresponding rotational period is shown, which indicates that the stars composed of hyperon rich matter can rotate much faster to be observed as a millisecond pulsar. For all the models presently studied, the neutron star seems to attain very high velocity in the range $\Omega_{k} = (6500 - 7500) s^{-1}$ at maximum mass, which corresponds to a kepler period of $P_k = (0.966 - 0.837)$. It is interesting to find that the stars corresponding to the softest two prescription i.e., G1 \& G2 attains minimal and maximal rotation respectively. Recently there has been considerable debate over the possibility of finding a millisecond pulsar and its implication on the star structure. In the Fig. 5(A), we plot the two fastest known pulsars till date and compare it with the results obtained in our calculations. The observation of rapid rotors such as XTE J1739-285  with spin frequency of $1122$ Hz is the most rapidly rotating neutron star yet discovered \cite{karat}. Prior to the discovery of 1122 Hz, the discovery of a radio pulsar spinning at 716 Hz suggests that the true maximum spin rate can be even higher \cite{hess06}. However, there are few more observations of pulsar period rotating in the vicinity, namely the $PSR B 1937 + 21$ ($\nu~=633~Hz$) \cite{633} and $PSR B 1957 + 20$ ($\nu~=621~Hz$) \cite{621}. It is noteworthy that the minimum observed pulsar period is still 1.56 ms and it is often argued that various mechanisms such as the r-mode instability and gravitational radiation may help in the slow down of the pulsar velocity and hence the millisecond pulsars are an exception. Such high rotations would invariably throw interesting possibilities pertaining to the structure and composition of a compact star, which we discuss next.

Rapid rotation can cause substantial deformation in the stellar structure as the matter is thrown in the equatorial plane, thereby resulting in flattening effect at the poles. On similar grounds, the decrease in the central density of the star can be understood in case of fast rotors. From the plot (Fig. 5(B)), we find that NL3, which results in the stiffest EoS among the five models studied here, undergoes maximum deformation and G2 is the least deformed star. However, for all the models the star seems to undergo maximal deformation at $\Omega_k~\approx~ 4700 s^{-1}$, ($f \approx 716 Hz$) thereafter the star rotates at much higher frequency without any further deformation (an increase in velocity of $\approx$ 50\%). The shape of the star for in the static limit and the corresponding star rotating with kepler velocity at maximum mass obtained, is shown in Fig. 6, for the five models in the polar-equatorial plane. We find that G2 sustains maximal keplerian velocity and the G1 is credited with the least, although the resulting deformation for the two remains the same. However, at maximum mass the star rotates with nearly equal frequency in case of TM1 and TM1*. As expected, the stiffness of the EoS showed up in case of NL3, which results in maximal deformation and G2 deforms the least. For a better comparison, the global properties of the neutron star with hyperon core, corresponding to the static and the kepler limit for the models under investigation are tabulated in Table II. 

\begin{table}
\caption{Comparison of the global properties of the neutron star in the static limit i.e., $\Omega = 0$ (upper row) and the stars rotating with kepler velocity ($\Omega = \Omega_k$) at maximum mass (lower row) for the corresponding EoS as tabulated.}
\begin{center}
\begin{tabular}{cccccccccccccc}
\hline
\hline
\multicolumn{1}{c}{EoS} &
\multicolumn{1}{c}{$M$} &
\multicolumn{1}{c}{$R_{eq}$} &
\multicolumn{1}{c}{$\varepsilon_{c}/\varepsilon_{0}$} &
\multicolumn{1}{c}{$M_{b}$} &
\multicolumn{1}{c}{$I_{45}$} &
\multicolumn{1}{c}{$f$} &
\multicolumn{1}{c}{$R_p / R_e$} \\
\hline
\multicolumn{1}{c}{} &
\multicolumn{1}{c}{($M_{\odot}$}) &
\multicolumn{1}{c}{}($Km$) &
\multicolumn{1}{c}{} &
\multicolumn{1}{c}{($M_{\odot}$)} &
\multicolumn{1}{c}{($10^{45} g cm^2$)}&
\multicolumn{1}{c}{($Hz$)} &
\multicolumn{1}{c}{}\\
\hline
\hline
NL3    &1.79   &13.20   &4.64  &2.00  &--   &--   &-- \\
       &2.18   &18.23   &4.40  &2.44  &3.92 &1095 &0.56 \\
\hline
TM1    &1.51   &12.36   &6.17  &1.65  &--   &--   &-- \\
       &1.80   &17.06   &6.00  &1.97  &2.49 &1100 &0.58\\
\hline
TM1*   &1.42   &11.30   &8.99  &1.55  &--   &--   &-- \\
       &1.68   &16.50   &6.59  &1.83  &2.11 &1117 &0.58\\
\hline
G1     &1.33   &12.45   &5.06  &1.44  &--   &--   &-- \\
       &1.58   &17.33   &5.00  &1.70  &2.05 &1009 &0.59\\
\hline
G2     &1.29   &10.64   &10.15 &1.40  &--   &--   &-- \\
       &1.48   &15.25   &8.42  &1.61  &1.47 &1183 &0.59\\
\hline
\end{tabular}
\end{center}
\end{table}

\section{CONCLUSIONS}

We study the effect of nuclear interaction terms on the equation of state of hyperon rich matter and the resulting consequence on the neutron star properties in both the static and the kepler limit. The effect of rapid rotation on the stellar structure is investigated with different models and the effect of rotation on the neutron star structure is lineated. Rotating neutron stars, can potentially not only constrain the EoS of dense matter, but is also interesting because they can be the source of gravitational wave radiation, the mechanism which is believed to take away the star angular momentum thereby resulting in the slow down of the stars. For the purpose, we took five different models from effective mean-field theory with different inherent features, such as the model interactions and the nuclear saturation properties to analyze their effect on the equation of state of dense matter. The inclusion of extra terms in the lagrangian contributes to the softening of the EoS and the matter becomes more compressible. From the result of the properties of the star, we found that a soft EoS (one with lower value of incompressibility) results in star with small maximum mass, small radius, and large rotation rate and vice versa. In comparison to the static stars, the rotating stars gain more mass. We found an increase of $\approx$ (15-20)\% in the maximum mass, with a corresponding decrease of $\approx$ (5-17) \% in the central density of the star rotating with kepler velocity. The models can also explain fast rotors, typically $f > 1 kHz$, which is in agreement with the fastest observed frequency of $f=1122 Hz$. However, it was rather interesting that the resulting deformation almost remains the same after it attains the next highest recorded frequency of $f=716 Hz$, i.e., the star seems to sustain faster rotations (an increase in rotational frequency up to $\approx$ 50\%) without any further deformation. Also, at kepler velocity, independent of the compressibility, all the models presently under consideration attains similar deformation in magnitude. The observation of millisecond pulsar would constrain the equation of state of matter at high densities, however, for various dynamical processes, such as the R-mode instability may prevent neutron stars from reaching such high spin rates \cite{rmode}. It would be interesting to study the aspects and effect of r-modes in regulating the spin of neutron stars within similar framework. Work is in progress in this direction.

\end{document}